\documentclass{article}

\usepackage{amsmath, amsthm, amssymb}
\usepackage{graphicx}
\usepackage{comment}

\newtheorem{theorem}{Theorem}
\newtheorem{proposition}[theorem]{Proposition}
\newtheorem{lemma}[theorem]{Lemma}

\theoremstyle{definition}
\newtheorem{definition}{Definition}

\newcommand{\pats}{{\scshape Pats}}
\newcommand{\sat}{{\scshape Sat}}

\newcommand{\NP}{{\tt NP}}

\newcommand{\north}{{\tt N}}
\newcommand{\west}{{\tt  W}}
\newcommand{\south}{{\tt S}}
\newcommand{\east}{{\tt  E}}

\newcommand{\xnor}{{\tt XNOR}}

\newcommand{\dom}{\ensuremath{\mathrm{dom}}}
\newcommand{\color}{\ensuremath{\mathrm{color}}}

\newcommand{\nc}{59}

\newcommand{\tas}{\ensuremath{\mathcal{T}}}

\def\vector#1{\mbox{\boldmath $#1$}}

\title{Combinatorial Optimization in Pattern Assembly}
\author{Shinnosuke Seki}

\begin{document}

\maketitle

\begin{abstract}
	Pattern self-assembly tile set synthesis ({\pats}) is a combinatorial optimization problem which aim at minimizing a rectilinear tile assembly system (RTAS) that uniquely self-assembles a given rectangular pattern, and is known to be \NP-hard. 
	{\pats} gets practically meaningful when it is parameterized by a constant $c$ such that any given pattern is guaranteed to contain at most $c$ colors ($c$-{\pats}). 
	We first investigate simple patterns and properties of minimum RTASs for them. 
	Then based on them, we design a {\nc}-colored pattern to which {3\sat} is reduced, and prove that {\nc}-{\pats} is \NP-hard. 
\end{abstract}

	\section{Introduction}

Tile self-assembly is an algorithmically rich model of ``programmable crystal growth.'' 
Well-designed molecules (square-like ``tiles'') with specific binding sites can deterministically form a single target shape even subject to the chaotic nature of molecules floating in a well-mixed chemical soup. 
Such tiles was experimentally implemented as DNA double-crossover molecules in 1998 \cite{WiLiWeSe1998}, and the last decade saw the drastic improvement of the reliability of DNA tile self-assembly; see, e.g., \cite{BaScRoWi2009}. 

Shape-building is one primary goal of self-assembly; pattern-painting is another. 
Based on the abstract Tile Assembly Model (aTAM) introduced by Winfree \cite{Winfree_PhDthesis}, Ma and Lombardi have first shed light on this problem and formalized it in the name of {\em patterned self-assembly tile set synthesis} ({\pats}) problem in \cite{MaLombardi2008,MaLombardi2009}.
In this framework, an optimization problem of interest aims at minimizing the number of tile types necessary for a rectilinear TAS (RTAS) to uniquely assemble a given rectangular pattern. 
An exhaustive partition-search algorithm \cite{GoosOrponenDNA16} as well as a randomized search algorithm \cite{LempiainenCzeizlerOrponenDNA17} have been proposed for this problem. 

{\pats} was recently proved to be \NP-hard\footnote{
	This problem had been claimed \NP-hard even with $d=2$ \cite{MaLombardi2009}, but an error was found as being pointed out in \cite{CzeizlerPopa2012}. 
}. 
Nevertheless, it is not until being investigated under the restriction of the number of available colors that {\pats} gets practically meaningful, as summarized in DNA 2012 as: ``{\it any given logic circuit can be formulated as a colored rectangular pattern with tiles, using only a constant number of colors}.'' 
We will hence propose a variant of {\pats} parameterized by the size of palette $c$ and propose {\em $c$-{\pats}}. 
The main contribution of this paper is the proof of the \NP-hardness of $c$-{\pats} for $c = \nc$.  

Various techniques that have been invented for combinatorial optimization in shape assembly (see e.g.,~\cite{AdChGoHu2001,AdChGoHuKeEsRo2002,BrChDoKaSe2011,ChenDotySeki2011}) are useful but not sufficient in the proof because we now encounter a new challenge intrinsic to the optimization in pattern assembly. 
That is the combinatorial explosion in the number of possible ways to color tile types. 
This is observed even in assembling a two-colored pattern by TASs. 
Two tile types are trivially necessary for that (there is no chameleon tile type). 
If a TAS can use only 2 tile types, then there is no choice but to draw the types by distinct colors. 
By contrast, once more tile types become available, TAS designers cannot do without considering how many tile types to be painted by a color\footnote{ 
It is not fair to say only due to this explosion that combinatorial optimization is more challenging for (multicolor) pattern assembly than shape building (monotone pattern assembly); colors often provide a visual clue to distinguish types of tiles. 
}.
Since there is a pattern whose tile complexity is greater than the number of colors it contains, this combinatorial explosion must be addressed in some way. 
In Section~\ref{subsec:basics}, we will propose subpatterns that are embedded into a bigger pattern $P$ and cooperatively force RTASs that uniquely assemble $P$ to paint at least certain number of their tile types with a specific color. 
Being combined with an upperbound on the number of tile types available for the RTASs, which is given in $c$-{\pats}, the subpatterns prove their worth of providing the RTASs with the precise number of tile types to be drawn by each color. 

	\section{Preliminaries}

In this section, we mainly recall the abstract Tile Assembly Model (aTAM) proposed by Winfree \cite{Winfree_PhDthesis} and the problem {\sc Pattern self-Assembly Tile set Synthesis} (\pats) proposed by Ma and Lombardi \cite{MaLombardi2008}. 
A variant of {\pats} will be the main focus of this paper. 

	\subsection{Abstract Tile Assembly Model (aTAM)}
	\label{subsec:def_aTAM}

Let $\Sigma$ be an alphabet, and by $\Sigma^*$, we denote the set of finite strings over $\Sigma$. 
By $\mathbb{Z}$ and $\mathbb{N}$, we denote the set of integers and the set of positive integers, respectively, and let $\mathbb{N}_0 = \mathbb{N} \cup \{0\}$. 
In aTAM, $\mathbb{Z}^2$ is especially considered either as the two-dimensional integer lattice or as the set of all points on it. 

Given a set of points $A \subseteq \mathbb{Z}^2$ on the integer lattice, the {\it full grid graph} of $A$ is the undirected graph $G^{\rm f}_A = (V, E)$, where $V = A$ and for all $u, v \in V$, there is an edge between $u$ and $v$ if and only if $||u-v||_2 = 1$, where $|| \cdot ||_2$ is the Manhattan distance, that is, $u$ and $v$ are adjacent points. 
Let $\north, \west, \south, \east$ stand for the respective directions north, west, south, and east, and be also interpreted as the respective unit vectors $(0, 1), (-1, 0), (0, -1), (1, 0)$. 

\begin{figure}[tb]
\begin{center}
\begin{minipage}{0.75\textwidth}
\includegraphics[scale=0.45]{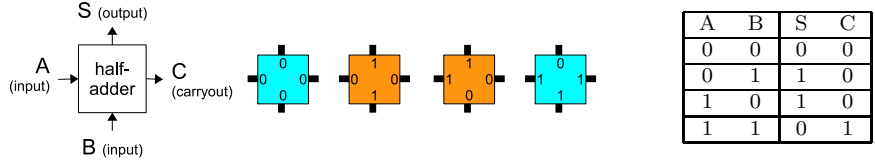}
\end{minipage}
\begin{minipage}{0.2\textwidth}
{\footnotesize
\begin{tabular}{|cc|cc|}
\hline
A & B & S & C \\
\hline
0 & 0 & 0 & 0 \\
\hline
0 & 1 & 1 & 0 \\
\hline
1 & 0 & 1 & 0 \\
\hline
1 & 1 & 0 & 1 \\
\hline
\end{tabular}
}
\end{minipage}
\end{center}
\caption{
Four tile types (two blues, two oranges) implement together the half-adder, with two inputs A, B from the west and south, the output S to the north, and the carryout C to the east. 
Just for reference, the truth table of half-adder is also presented. 
}
\label{fig:half-adder}
\end{figure}

A {\it tile type} $t$ is a quadruple $t \in \Sigma^* \times \Sigma^* \times \Sigma^* \times \Sigma^*$, and is regarded as a unit square with four sides listed in the counter-clockwise order starting at the north (\north), each having a {\it glue label} (a.k.a., {\it glue}) taken from $\Sigma^*$; for instance, the second rightmost (orange) tile type in Figure~\ref{fig:half-adder} is represented as $(1, 1, 0, 0)$. 
For each direction $d \in \{\north, \west, \south, \east\}$, let $t(d)$ be the glue label at the $d$ side of $t$. 
Let $T$ be a {\it finite} set of tile types, and let us denote the (finite) set of all glues of tile types in $T$ by $\Lambda(T) \subseteq \Sigma^*$. 
An {\it assembly} (a.k.a., {\it supertile}) is a positioning of tiles of types in $T$ on (part of) the integer lattice $\mathbb{Z}^2$.  
It does not have to be a tessellation. 
Hence, we can say that an assembly is a partial function $\mathbb{Z}^2 \dashrightarrow T$.
Given two assemblies $\alpha, \beta: \mathbb{Z}^2 \dashrightarrow T$, $\alpha$ is a {\it sub-assembly} of $\beta$, written as $\alpha \sqsubseteq \beta$, if $\dom(\alpha) \subseteq \dom(\beta)$ and for every point $p \in \dom(\alpha)$, $\alpha(p) = \beta(p)$, where $\dom$ denotes the domain of the function. 

The aTAM models dynamics in the growth of assemblies based on the interaction among its basic building blocks, tiles. 
A {\it strength function $g: \Lambda(T) \to \mathbb{N}_0$} endows tiles with an ability to interact with its neighboring tiles by assigning the strength $g(\ell)$ to the {\it matching} label $\ell$ of their abutting edges.
If the labels do not match or $g(\ell) = 0$, these tiles do not interact; otherwise, they do. 
Tile interactions according to $g$ have an assembly $\alpha$ induce a {\it binding graph}, which is a grid graph whose vertices are $\dom(\alpha)$ and for two neighboring positions $p_1, p_2 \in \dom(\alpha)$, there is an edge between $p_1$ and $p_2$ on this graph if and only if the tiles $\alpha(p_1)$ and $\alpha(p_2)$ interact.  
On this graph, an edge between vertices means that the corresponding tiles interact, and hence, their abutting edges share the same label $\ell$. 
Thus, we can consider that the edge is labeled with $\ell$ and $g$ gives it the weight $g(\ell)$. 
The assembly is {\it $\tau$-stable (with respect to $g$)} if every cut of its binding graph has strength at least $\tau$. 
That is, the assembly is $\tau$-stable if at least energy $\tau$ is required to separate it into two parts. 

A {\it (seeded) tile assembly system} (TAS) is a quadruple $\tas = (T, \sigma, g, \tau)$, where $T$ and $g$ are as stated above, $\tau \ge 1$ is an integer parameter called {\it temperature}, and $\sigma$ is a finite $\tau$-stable {\it seed assembly} consisting of tile types that are NOT included in $T$. 
$\tas$ is provided with inexhaustible supply of copies of each tile type in $T$, each copy being referred to as a {\it tile}. 

Given two $\tau$-stable assemblies $\alpha, \beta$, we write $\alpha \to_1^{\tas} \beta$ if $\alpha \sqsubseteq \beta$, $\dom(\beta) \setminus \dom(\alpha) = \{p\}$ for some position $p \in \mathbb{Z}^2$, and $\beta(p) \in T$. 
Intuitively, this means that $\alpha$ can grow into $\beta$ by the addition of a single tile in $T$ at the position $p$. 
Since $\beta$ is required to be $\tau$-stable, the new tile binds to $\alpha$ with strength at least $\tau$. 
In this case, we say that {\it $\alpha$ $\tas$-produces $\beta$ in one step}. 

A sequence of $\tau$-stable assemblies $\alpha_0, \alpha_1, \ldots, \alpha_k$ is a {\it $\tas$-assembly sequence} if for all $1 \le i \le k$, $\alpha_{i-1} \to_1^{\tas} \alpha_i$ holds. 
We write $\alpha \to^\tas \beta$ and say {\it $\alpha$ $\tas$-produces $\beta$} (in 0 or more steps) if there is a $\tas$-assembly sequence $\alpha_0, \alpha_1, \ldots, \alpha_k$ of length $k = |\dom(\beta) \setminus \dom(\alpha)|$ with $\alpha_0 = \alpha$ and $\alpha_k = \beta$\footnote{
	This definition of producibility is justified by our limited focus only onto the finite assemblies in this paper; for the infinite assembly, it is not appropriate; see \cite{BrChDoKaSe2011} for instance.
}.
An assembly $\alpha$ is {\it $\tas$-producible} or {\it producible by $\tas$} if $\sigma \to^\tas \alpha$. 
A $\tau$-stable assembly $\alpha$ is {\it ($\tas$-)terminal} if for any $\tau$-stable assembly $\beta$, $\alpha \to^\tas \beta$ implies $\alpha = \beta$. 
Let $\mathcal{A}[\tas]$ be the set of assemblies producible by $\tas$, and let $\mathcal{A}_\Box[\tas] \subseteq \mathcal{A}[\tas]$ be the set of terminal assemblies that are producible by $\tas$. 
A TAS $\tas$ is {\it directed} if for each $\alpha, \beta \in \mathcal{A}[\tas]$, there exists $\gamma \in \mathcal{A}[\tas]$ such that $\alpha \to^\tas \gamma$ and $\beta \to^\tas \gamma$. 
One can easily verify that $\tas$ is directed if and only if $|\mathcal{A}_\Box[\tas]|=1$, and that $\tas$ is {\em not} directed if and only if there exist $\alpha, \beta \in \mathcal{A}[\tas]$ and a position $p \in \dom(\alpha) \cap \dom(\beta)$ such that $\alpha(p) \neq \beta(p)$.  

Throughout this paper, any TAS $\tas = (T, \sigma, g, \tau)$ is assumed to be free from useless tile types in the sense that for any tile type $t \in T$, there are a $\tas$-producible assembly $\alpha \in \mathcal{A}[\tas]$ and a position $p \in \dom(\alpha)$ such that $\alpha(p) = t$. 

	\subsection{Rectangular patterns and rectilinear TASs}
	\label{subsec:RTAS}

For $w, h \ge 1$, a {\it (rectangular) pattern (of width $w$ and height $h$)} is a function from the rectangular domain $\{(x, y) \mid x \in \{0, 1, \ldots, w\}, y \in \{0, 1, \ldots, h\}\}$ to $\mathbb{N}$. 
We call the image of a pattern $P$ the {\em color set of $P$}, and denote it by $\color(P)$. 
That is, any color in $\color(P)$ occurs at least once on $P$. 
When $|\color(P)| \le k$, we say that $P$ is {\it $k$-colored}. 

The {\it rectilinear TAS} (RTAS) is a variant of {\em temperature-2} TAS that is specialized for the rectangular pattern assembly. 
An RTAS is a quadruple $\tas = (T, \sigma, g, 2)$, where  
\begin{enumerate}
\item	$T$ and $g$ are as defined for TASs; 
\item	$\sigma$ is a 2-stable seed of L-shape, whose domain is $\{(x, 0) \mid 0 \le x \le w\} \cup \{(0, y) \mid 0 \le y \le h\}$; 
\item	for any $t \in T$ and $d \in \{\north, \west, \south, \east\}$, $g(t(d)) = 1$. 
\end{enumerate}
The statement 3 characterizes the rectilinear manner of assembly. 
The temperature being 2, the statement implies that any tile attachment needs the cooperation between the west glue and south glue, both of which are of strength~1.   
As such, at the initial time point of assembly process, for example, the sole position at which a tile in $T$ can attach is $(1, 1)$. 
Then it is a routine to verify that the assembly process proceeds from south-west to north-east {\it rectilinearly}. 
We say that $\tas$ {\it uniquely self-assembles a $k$-colored pattern $P$} if there exists a coloring function $f$ from $T \cup T_\sigma$ to the color set $C = \{1, 2, \ldots, k\}$ such that for any terminal assembly $\alpha \in \mathcal{A}_\Box[\tas]$, $f(\alpha) = P$.

\begin{figure}[tb]
\begin{center}
\includegraphics[scale=0.3]{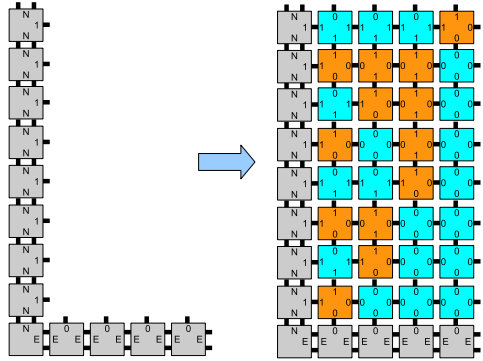}
\end{center}
\caption{
	From the L-shape seed, an RTAS $\tas_{\rm bc}$ uniquely self-assembles the infinite binary counter using the four tile types that implement the half-adder. 
}
\label{fig:SA_counter}
\end{figure}

As an example, an RTAS $\tas_{\rm bc}$ that uniquely self-assembles the infinite binary counter is shown in Figures~\ref{fig:half-adder} and~\ref{fig:SA_counter}. 
Two blue tile types and two orange ones in Figure~\ref{fig:half-adder} represent all four possiblities of 2-bit inputs {\tt 00}, {\tt 01}, {\tt 10}, {\tt 11} as their west and south glues, output the sum of the inputs to the north, and carry out to the east. 
In this way, they implement the half-adder together. 
$\tas_{\rm bc}$ fills the first quadrant defined by its L-shape seed by these four tile types in the {\em directed} manner (Figure~\ref{fig:SA_counter}). 

\begin{figure}[tb]
\begin{center}
\includegraphics[scale=0.45]{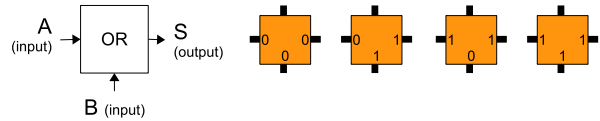}
\end{center}
\caption{
Four tile types implement together the {\tt OR}-gate, with two inputs A, B and one output S. 
}
\label{fig:or-gate}
\end{figure}

The L-shape seed of $\tas_{\rm bc}$ provides the sequence of 1-glues to the east and that of 0-glues to the north. 
With different glue sequences on the seed, tiles of $\tas_{\rm bc}$ uniquely yield other blue-orange patterns. 
Thus, we can say that the tile type set of $\tas_{\rm bc}$ is a machanism to convert a glue sequence on the seed given as input to a rectangular pattern. 
Furthermore, this example suggests that not only the half-adder but any combinatorial logics with at most two inputs and at most two outputs can be thus implemented using at most four tile types. 
Figure~\ref{fig:or-gate} presents such an implementation of {\tt OR}-gate, for instance. 
In Section~\ref{sec:reduction}, we shall design a set of tile types that evaluates a {3\sat} instance according to a given Boolean value assignment.

	\subsection{PATS: pattern self-assembly tile set synthesis problem}

Now, we introduce the pattern self-assembly tile set synthesis problem ({\pats}) originally proposed by Ma and Lombardi \cite{MaLombardi2008}. 
{\pats} aims at computing the minimum size RTAS that uniquely self-assembles a given rectangular pattern, where the size of an RTAS is measured by the cardinality of its tile type set. 

\begin{definition}[{\scshape Pattern self-Assembly Tile set Synthesis (Pats)}] 
\ \\
\begin{tabular}{ll}
{\scshape Given}: & a pattern $P$ \\
{\scshape Find}:  & a smallest directed RTAS that uniquely self-assembles $P$. 
\end{tabular}
\end{definition}

\noindent
Note that the solution to {\pats} is required to be directed here, while it was not so in its original definition. 
This, however, does not change the problem, as being observed in \cite{GoosOrponenDNA16}. 

{\pats} is an \NP-hard problem \cite{CzeizlerPopa2012}. 
By parameterizing it by the number of maximum colors $c$ used to draw patterns, we propose the following more practically meaningful variant of {\pats}. 

\begin{definition}[{\scshape $c$-colored Pattern self-Assembly Tile set Synthesis ($c$-Pats)}] 
\ \\
\begin{tabular}{ll}
{\scshape Given}: & a $c$-colored pattern $P$; \\
{\scshape Find}:  & a smallest directed RTAS that uniquely self-assembles $P$. 
\end{tabular}
\end{definition}

We will find out a constant $c$ that makes {$c$-\pats} \NP-hard, through a polynomial-time reduction of {3\sat} to the following decision variant of the problem: 

\vspace*{2mm}
\begin{tabular}{ll}
{\scshape Given}: 	& a $c$-colored pattern $P$ and an integer $n$; \\
{\scshape Output}:  	& {\tt YES} if there exists a directed RTAS with at most $n$ tile types \\ 
			& that uniquely self-assembles $P$. 
\end{tabular}

	\subsection{Basic Combinatorial Results}
	\label{subsec:basics}

Before proceeding to the main result, let us present several basic results on directed RTASs, which we will be used in Section~\ref{sec:reduction}. 
Let us begin with the most important property which characterize the directedness property by the west and south glues of their tile types. 

\begin{proposition}\label{prop:directed_RTAS_characterization}
	An RTAS is directed if and only if it contains no distinct tile types $t_1, t_2$ with $t_1(\west) = t_2(\west)$ and $t_1(\south) = t_2(\south)$\footnote{
		This proposition is true as long as all tile types of $\tas$ appear on some assembly producible by $\tas$. 
		This is an assumption we make in this paper, as declared at the end of Section~\ref{subsec:def_aTAM}. 
	}. 
\end{proposition}

This proposition enables us to design simple patterns that, being embedded into another pattern $P$ as a subpattern, necessitates {\em at least} 2 tile types in order for {\em any} directed RTAS to uniquely self-assemble $P$. 
One of such patterns can be found in the binary counter pattern $P_{\rm bc}$ in Figure~\ref{fig:SA_counter}. 
At the orange position $P_{\rm bc}(2, 2)$ and blue position $P_{\rm bc}(4, 2)$, an RTAS $\tas$ must put tiles of distinct types $t_1, t_2$. 
Moreover, in order for $\tas$ to be directed, $t_1$ needs to disagree to $t_2$ with respect to either west glue or south glue (Proposition~\ref{prop:directed_RTAS_characterization}). 
This implies that $\tas$ has two distinct blue tile types because the west and south neighbors of these positions are all blue. 
This observation is formally described as follows.

\begin{lemma}\label{lem:at_least_2-1}
	Let $\tas$ be a directed RTAS that uniquely self-assembles a pattern $P$. 
	For a color $i$ and positions $(x_1, y_1), (x_2, y_2)$, if $P(x_1-1, y_1) = P(x_1, y_1-1) = P(x_2-1, y_2) = P(x_2, y_2-1)=i$ but $P(x_1, y_1) \neq P(x_2, y_2)$, then $\tas$ has at least two tile types of color $i$. 
\end{lemma}

\begin{figure}[tb]
\begin{minipage}{0.6\textwidth}
\begin{center}
\includegraphics[scale=0.4]{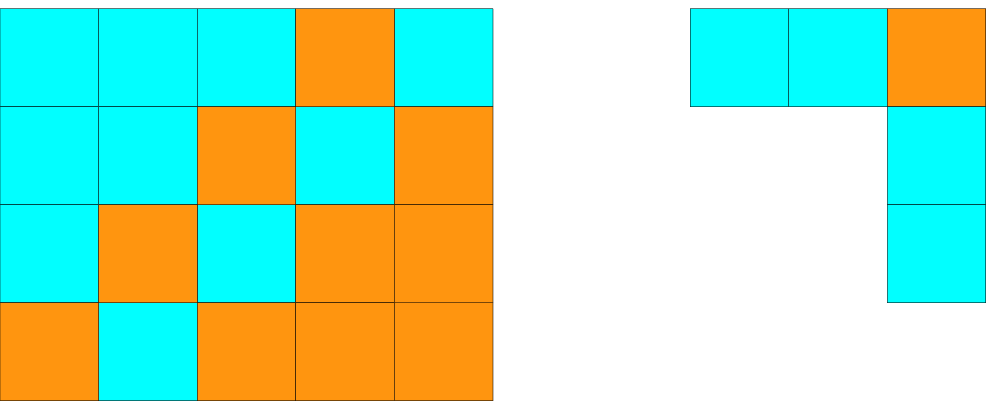}
\end{center}
\end{minipage}
\begin{minipage}{0.3\textwidth}
\includegraphics[scale=0.4]{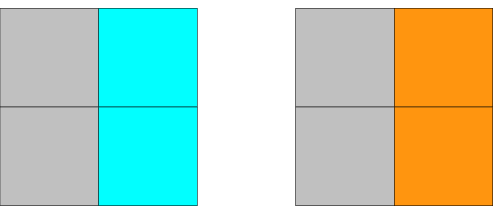}
\end{minipage}
\caption{
(Left) 	If this blue-orange binary-color subpattern appears on a pattern $P$, a directed RTAS needs at least 2 blue tile types and at least 2 orange tile types in order to uniquely self-assemble $P$; 
(Middle)	This pattern also asks a directed RTAS to prepare 2 blue tile types. 
(Right)
}
\label{fig:LB2}
\end{figure}

Lemma~\ref{lem:at_least_2-1} immediately leads us to the next lemma.

\begin{lemma}\label{lem:at_least_2-substitutor}
	If a directed RTAS uniquely self-assembles a pattern on which the pattern shown in Figure~\ref{fig:LB2} (Left) appears, then it has at least 2 blue tile types and at least 2 orange tile types. 
\end{lemma}

Another pattern of interest is shown in Figure~\ref{fig:LB2} (Middle). 
With tiles of type $t$ at all the four blue positions, $t$ would satisfy $t(\west) = t(\east)$ and $t(\north) = t(\south)$ and hence a blue tile (of this type) would fill the orange position. 
This observation is formally described as follows.

\begin{lemma}\label{lem:at_least_2-2}
	Let $\tas$ be a directed RTAS that uniquely self-assembles a pattern $P$. 
	For a color $i$ and a position $(x, y)$, if $P(x-2, y) = P(x-1, y) = P(x, y-1) = P(x, y-2-1)=i$ but $P(x, y) \neq i$, then $\tas$ has at least two tile types of color $i$. 
\end{lemma}

\begin{figure}[tb]
\begin{center}
\includegraphics[scale=0.6]{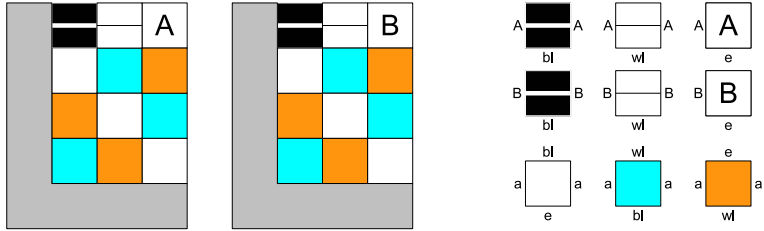}
\end{center}
\caption{
Two subpatterns that cooperatively force at least 2 colors to be used to draw at least 2 tile types.
To the right, a set of 9 tile types that uniquely self-assemble them is presented, which is the sole minimum one. 
}
\label{fig:LB2_multiple}
\end{figure}

Next, we propose a mosaic pattern {\em parameterized by an integer $k$} that forces $k$ number of different colors to be used to draw at least 2 tile types. 
For $k = 2$, this pattern can be found in Figure~\ref{fig:LB2_multiple}.
It contains 7 colors: white, blue, orange, lined black, lined white, {\tt A}, and {\tt B}. 
On the assumption that the gray part be the seed, where, by definition, any information can be encoded, how many tile types are necessary and sufficient for a directed RTAS $\tas$ to uniquely self-assemble them. 
Since ${\tt A} \neq {\tt B}$, the two tile types $t_A, t_B$ need to satisfy either $t_A(\west) \neq t_B(\west)$ or $t_A(\south) \neq t_B(\south)$. 
In the former case, as shown in Figure~\ref{fig:LB2_multiple}, two lined black types and two lined white ones can carry the one bit information (to be {\tt A} or not to be), and one white, blue, and orange are enough. 
On the other hand, in the latter case, we need to deliver the 1-bit information through the 3-colors (white, blue, orange) mosaic pattern and no matter how it is delivered, the path encounters all of these 3 colors. 
If it were not for two blue tile types, the assembly process would lose the information once it hits the diagonal blue stripe, and this argument is valid for white and orange. 
Thus in this case we need extra 3 tile types in contrast to the need for 2 tile types in the former case. 
So, for instance, with only 9 tile types, the latter propagation strategy cannot be employed.  
Let us note that this parameterized mosaic pattern is ``stretchable'' by duplicating each column arbitrary many times.  

We have seen several patterns that forces an RTAS to draw 2 tile types by a specific color. 
It is a challenging issue to design a pattern {\em without introducing many auxiliary (wasteful) colors} that forces more than 2 types to be drawn by a specific color. 
Another issue of significance is to design a pattern that allocates glues on tile types in an intended manner. 
In Figure~\ref{fig:LB2} (Right), you see two $2 \times 2$ square patterns, which consist of three colors (gray, blue, orange). 
If a directed RTAS uniquely self-assembles a pattern including these square patterns, and moreover, using {\em only one} tile type of each of these three\footnote{In fact, it is fine that gray is equal to either blue or orange as long as blue is distinct from orange in order to reach the coming conclusion.} colors. 
Let them be $t_g, t_b, t_o$. 
Then $t_g(\east) = t_b(\west) = t_o(\west)$. 
Thus, $t_b(\south) \neq t_o(\south)$ must hold (Proposition~\ref{prop:directed_RTAS_characterization}). 
This can be strengthened as $t_b(\north) = t_b(\south) \neq t_o(\south) = t_o(\north)$ because blue and orange positions are in tandem.

	\section{Polynomial-Time Reduction of 3SAT to \nc-Colored PATS}
	\label{sec:reduction}

The following is the main theorem of this paper. 

\begin{theorem}\label{thm:cpats_NPhard}
	{\nc-\pats} is \NP-hard. 
\end{theorem}

\begin{figure}[tb]
\begin{center}
\includegraphics[scale=0.7]{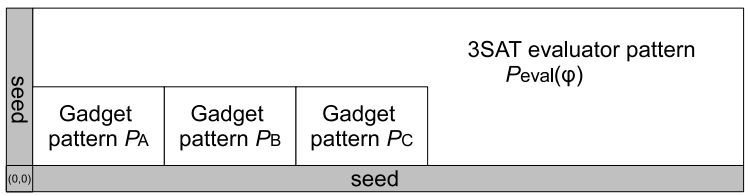}
\end{center}
\caption{The pattern $P(\phi)$ to which a {3\sat} instance $\phi$ is reduced.}
\label{fig:blueprint}
\end{figure}

Our proof takes the classic approach: a polynomial-time many-one reduction from {3\sat} to the decision variant of {$c$-\pats}. 
An instance of {3\sat} is a formula $\phi$ that is a conjunction of clauses consisting of exactly three literals (a variable or its negation); the $m$ variables of $\phi$ are indexed as $v_1$, $v_2$, and so on. 
We will propose a set $T_{\rm 3SAT}$ with ${\nc}+24$ tile types and a pattern $P(\phi)$ to which a given {3\sat} instance $\phi$ is reduced such that 
\begin{enumerate}
\item	any RTAS with this tile type set is directed due to Proposition~\ref{prop:directed_RTAS_characterization}; 
\item	if $\phi$ is satisfiable, a directed RTAS uniquely self-assembles $P(\phi)$ using $T_{\rm 3SAT}$; 
\item	if a directed RTAS uniquely self-assembles $P(\phi)$ using {\nc}+24 tile types, then its tile type set is isomorphic to $T_{\rm 3SAT}$ (up to glue label renaming) and (hence) $\phi$ is satisfiable. 
\end{enumerate}

As shown in Figure~\ref{fig:blueprint}, $P(\phi)$ consists of four subpatterns: {3\sat} evaluator pattern $P_{\rm eval}(\phi)$ and three gadget patterns $P_A, P_B, P_D$. 
The {3\sat} evaluator pattern is the main pattern and the other three gadget patterns play an auxiliary role in the proof of the third statement above. 

	\subsection{3SAT Evaluator Pattern}

\begin{figure}[tb]
\begin{center}
\includegraphics[scale=0.4]{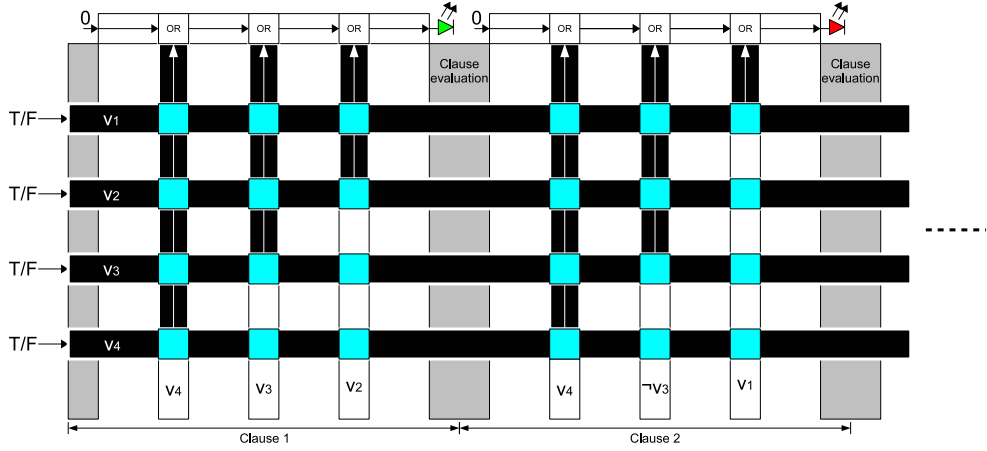}
\end{center}
\caption{
	A digital circuit to evaluate a specific {3\sat} instance $(v_4 \vee v_3 \vee v_2) \wedge (v_4 \vee \neg v_3 \vee v_1)$ according to a given assignment of Boolean values.
	Literals (white stripes) are evaluated at substitutors (blue squares), and the evaluation (black stripes with a white arrow) is transmitted to the north for the evaluation of clause they belong to. 
	It should be clear that the circuit design principle works for arbitrary number of variables or clauses. 
}
\label{fig:3sat_evaluator}
\end{figure}

Let us begin with the {3\sat} evaluator pattern. 
It is degisned based on a digital circuit which we call a {\it {3\sat} evaluator}. 
Just as the name suggests, it evaluates a given {3\sat} formula $\phi$ according to an assignment given as input. 
It must be noted first that this circuit is designed to be planar (no wire goes over the others) and rectilinear (signals always transmit from south-west to north-east) so that it can be easily transformed into $P(\phi)$. 

Figure~\ref{fig:3sat_evaluator} illustrates the {3\sat} evaluator. 
As input, it takes from the left an assignment $\vector{b} = (b_1, \ldots, b_m)$ according to which $\phi$ is evaluated, where $b_1, \ldots, b_m \in \{0, 1\}$ (0:false, 1:true). 
It is provided with $m$ horizontal wires, which transmit the input to the right along with the index $i$ of the variable they represent. 
It is also provided with vertical wires, which represent literals in $\phi$. 
We refer to the horizontal wire for variable $v_i$ simply as the {\it variable wire $v_i$}, and a vertical wire for literal $v_j$ (resp.~$\neg v_j$) as a {\it literal wire $v_j$ (resp.~$\neg v_j$)}. 
Any variable wire is connected with any literal wire at their intersection by a device called {\it substitutors}. 
At such an intersection between the variable wire $v_i$ and a literal wire $v_j$ or $\neg v_j$, the substitutor compares their indices, and if $i = j$, it substitutes the Boolean value $b_i$ into the literal by an {\xnor} gate and transmits the result to the north; otherwise, it does nothing but merely ``lets one go over the other.'' 
For each clause, its three literals thus evaluated are conjugated by three {\tt OR} gates, which determines whether the clause is satisfied or not, and the evaluation is output at the top as LED signal (red:unsatisfied, green:satisfied; see Figure~\ref{fig:3sat_evaluator}, in which Clause 1 is evaluated to be satisfied, whereas Clause 2 is not). 

\begin{figure}[p]
\begin{center}
\includegraphics[scale=0.75]{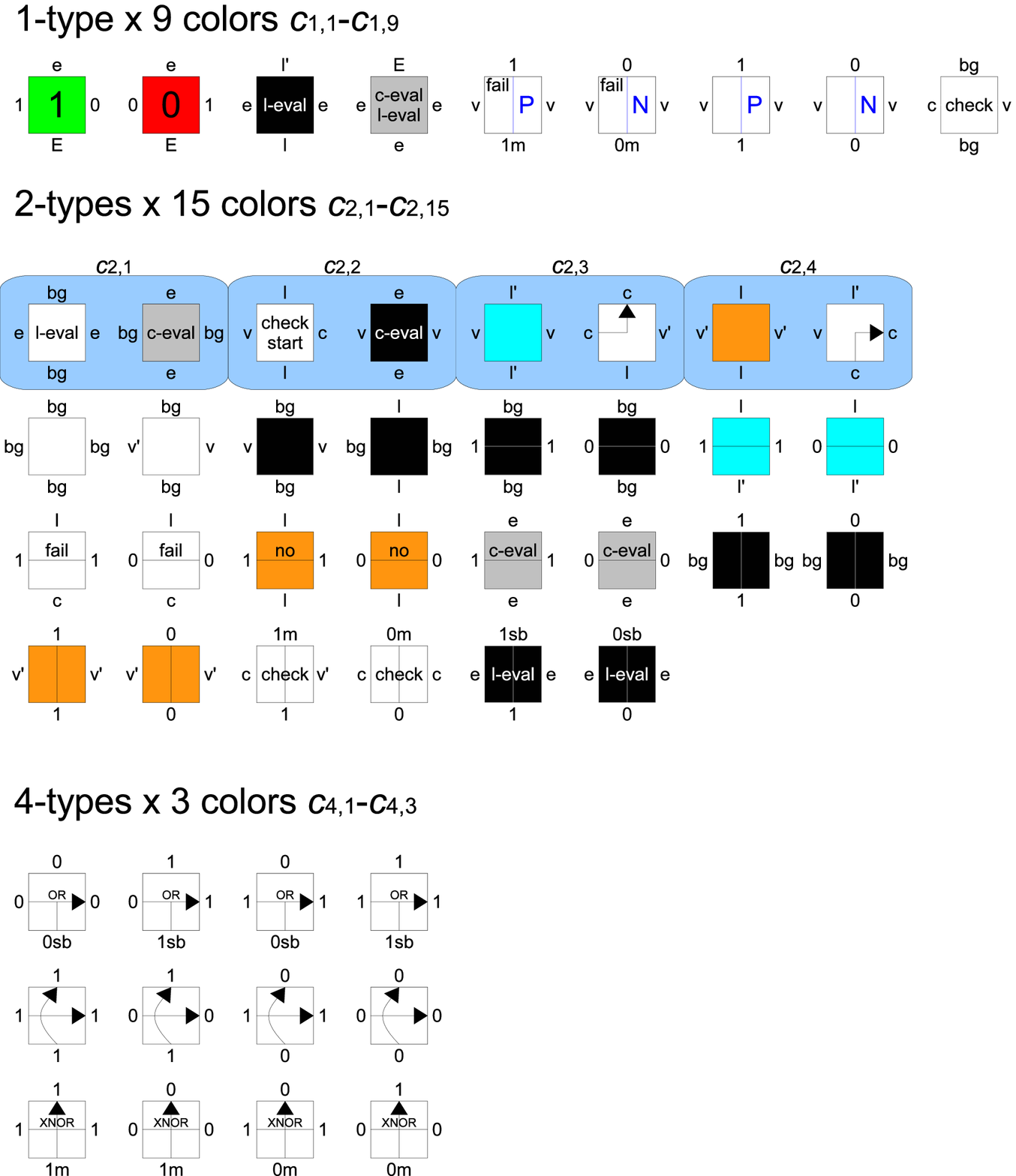}
\end{center}
\caption{
	The 51 tile types (of 27 colors) for the simulation of {3\sat} evaluator by a directed RTAS.
}
\label{fig:tileset}
\end{figure}

\begin{figure}[p]
\begin{center}
\includegraphics[scale=0.45]{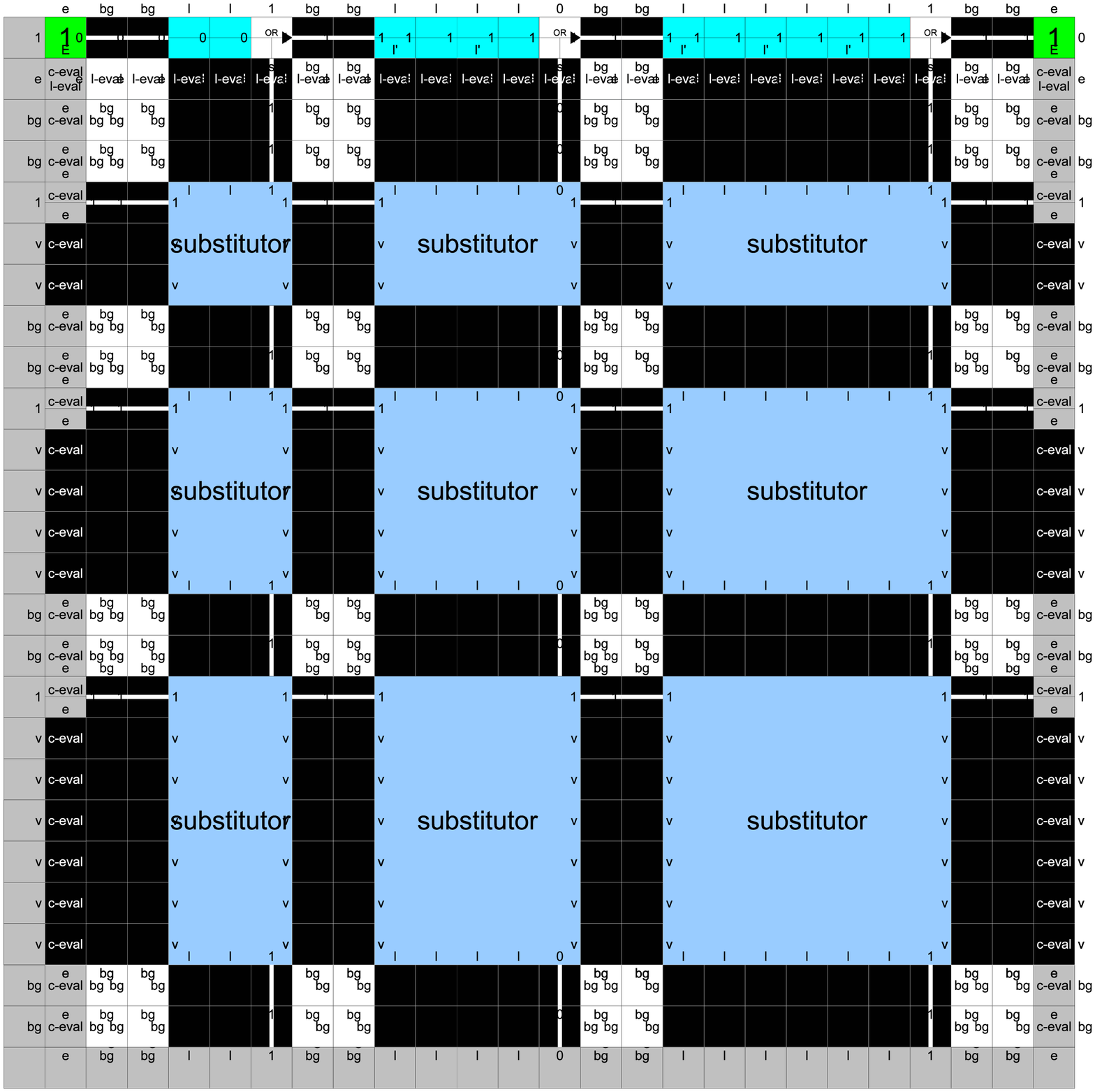}
\end{center}
\caption{Using tiles in the set $T_{\rm 3SAT}$, this pattern uniquely self-assembles from the L-shape seed that encodes a {3\sat} instance with 3 variables $v_1, v_2, v_3$ and a clause $\{v_1, \neg v_2, v_3\}$ with the assignment $(1, 1, 1)$.}
\label{fig:main_idea}
\end{figure}

The pattern $P_{\rm eval}(\phi)$ is a slight modification of a pattern $P_{\rm eval}(\phi, \vector{b})$ of {3\sat} evaluator's circuit layout. 
Now we will propose a tile type set $T_{\rm eval}$ (Figure~\ref{fig:tileset}) with which a directed RTAS simulates the {3\sat} evaluator and uniquely self-assembles $P_{\rm eval}(\phi, \vector{b})$. 
This consist of 51 tile types of 27 colors: 1 type of each of 9 colors $c_{1, 1}, c_{1, 2}, \ldots, c_{1, 9}$, 2 types of each of 15 colors $c_{2, 1}, \ldots, c_{2, 15}$, and 4 types of each of 3 colors $c_{4, 1}, c_{4, 2}, c_{4, 3}$. 
The assignment $\vector{b} = (b_1, \ldots, b_m)$ and instance $\phi$ are encoded on the L-shape seed. 
Specifically, the assignment $\vector{b}$ is encoded as the following glue sequence on the vertical bar of the seed: 
\begin{equation}\label{eq:assign_encoding}
	0 {\tt e} ({\tt bg})^2 \ b_1 {\tt v}^2 \ ({\tt bg})^2 \ b_2 {\tt v}^4 \ ({\tt bg})^2 \cdots ({\tt bg})^2 \ b_m {\tt v}^{2m} \ ({\tt bg})^2 @, 
\end{equation}
where $v_i$ is encoded with its assigned value $b_i$ as $b_i {\tt v}^{2i}$ and @ indicates the origin (0, 0), and ${\tt bg}, {\tt v}, {\tt e}$ are glues that stand for {\it background}, {\it variable}, and {\it evaluation}. 
On the other hand, the literal $v_j$ and its negation are encoded as ${\tt v}^{2j} 1$ and ${\tt v}^{2j} 0$, respectively (1:positive, 0:negative). 
Then we encode a clause $c = \{v_i, \neg v_j, v_k\}$ as: 
\begin{equation}\label{eq:literal_encoding}
	gs(c) = ({\tt bg})^2 \ {\tt l}^{2i} 1 \ ({\tt bg})^2 \ {\tt l}^{2j} 0 \ ({\tt bg})^2 \ {\tt l}^{2k} 1 \ ({\tt bg})^2 {\tt e},  
\end{equation}
where {\tt l} represents the {\it literal}. 
A glue sequence on the seed's horizontal bar that encodes the clauses $c_1, c_2, \ldots$ of $\phi$ is obtained by catenating $g(c_1)$, $g(c_2)$, and so on. 
Concatenating it further with the glue sequence in \eqref{eq:assign_encoding} amounts to the L-shape seed, from which, the pattern $P_{\rm eval}(\phi, \vector{b})$ self-assembles (see Figure~\ref{fig:main_idea} and compare it with the {3\sat} evaluator). 

\begin{figure}
\begin{center}
\includegraphics[scale=0.7]{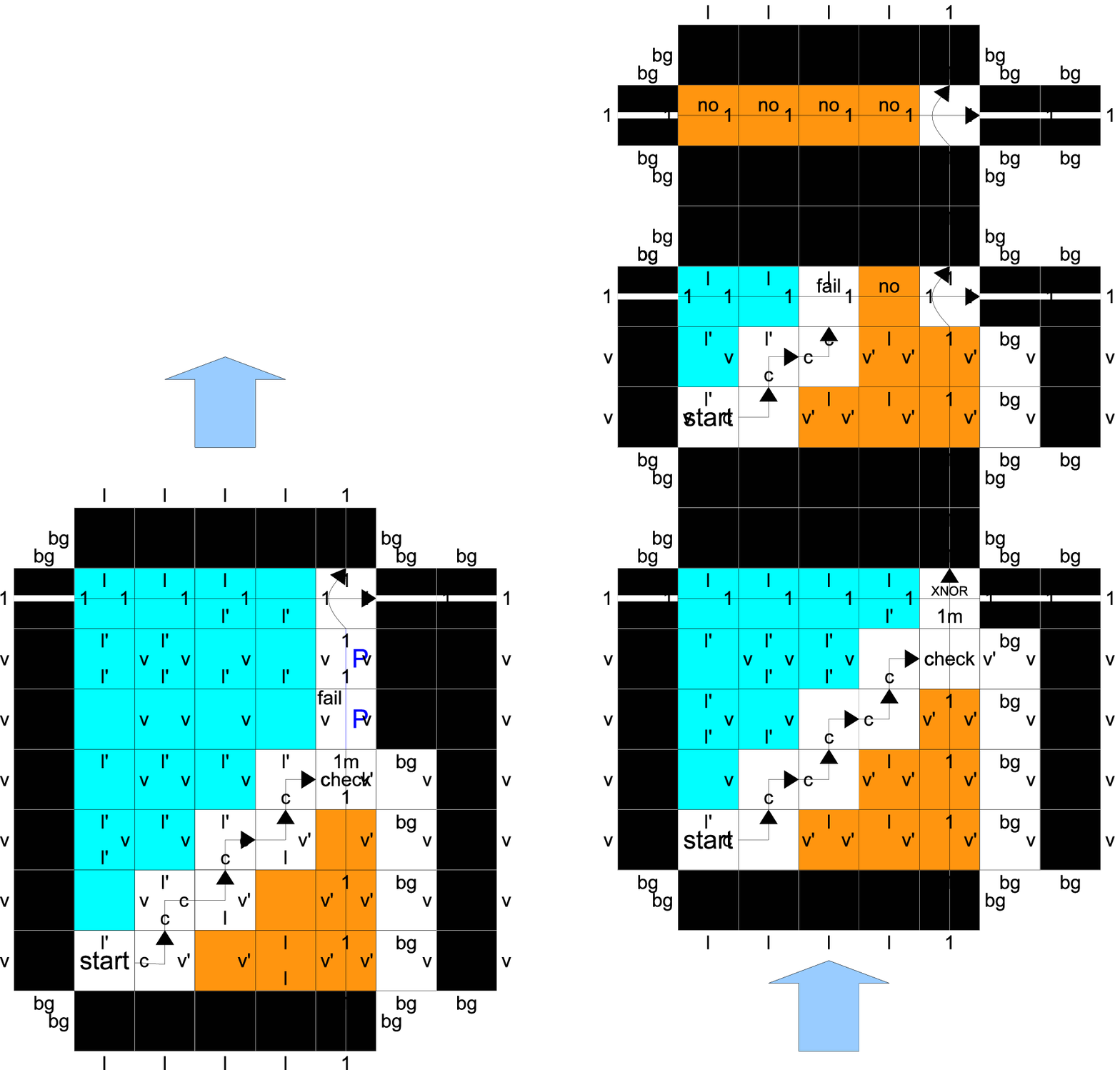}
\end{center}
\caption{
Patterns occurring at the intersection of the literal wire $v_2$ 
(Left) with the variable wire $v_3$, 
(Right bottom) with the wire for matching variable $v_2$, or
(Right, middle) with the variable wire $v_1$.
(Right, top) Its intersection with the thinnest wire, which does not encode any variable.
}
\label{fig:substitutors1}
\end{figure}

On $P_{\rm eval}(\phi, \vector{b})$, any literal wire encounters the $m$ variable wires. 
At the intersection of a literal wire $v_j$ with the variable wire $v_i$, tiles in $T_{\rm eval}$ self-assembles a pattern that visualizes the mechanism of substitutor in the {3\sat} evaluator. 
How this assembly proceeds may be understood more easily visually than with words, so see Figure~\ref{fig:substitutors1}. 
When the wires meet, the {\tt check-start} tile attachs and triggers the assembly of diagonal zig-zag snake. 
When it hits the literal wire, one of the 2 {\tt check} tile attachs, and if to its north, the variable wire $v_i$ is waiting with the input $b_i$ (this happens if and only if $i = j$), one of the 4 {\xnor} tiles selectively attachs, substitutes $b_i$ into the literal, and outputs the result to the north (1:true, 0:false; note that before and after the substitution, the signal 1/0 through a literal wire is interpreted in different ways). 
Figure~\ref{fig:substitutors1} shows how the literal wire goes over the variable wire without being substituted in other two cases when $i > j$ (Left) and $i < j$ (Right, middle). 
Due to the encoding \eqref{eq:assign_encoding}, at of the substitution, the wire $v_j$ has already crossed the variable wires $v_m, v_{m-1}, \ldots, v_{j+1}$, which are thicker than itself, and it is going to transmit the substituted value while crossing the remaining variable wires $v_{j-1}, \ldots, v_1$ in this order, which are thinner. 
With the property of $T_{\rm eval}$ that when a literal wire crosses a thicker variable wire, some tile types {\em visually} tell which of 1/0 the wire carries, while its crossing a thinner one leaves no such visual clue, the encounter order means that from $P_{\rm eval}(\phi, \vector{b})$, we cannot get any further clue of $\vector{b}$ than whether clauses are satisfied or not. 
This cover-up plays a critical role in the proof of \NP-hardness. 

The evaluation of each clause by 4 tile types of {\tt OR} color should be straightforward from Figure~\ref{fig:main_idea} (see its top).

The {\em only} positions on $P(\phi, \vector{b})$ whose color changes depending on the encoded assignment $\vector{b}$ are the LED positions. 
Drawing all these positions by green (satisfied) yields the pattern $P_{\rm eval}(\phi)$. 
By this construction, it is obvious that for a satisfiable $\phi$, a directed RTAS uniquely self-assembles $P_{\rm eval}(\phi)$, and actually the whole pattern $P(\phi)$. 
Although for an unsatisfiable $\phi$, $P(\phi)$ cannot be uniquely self-assembled by $T_{\rm 3SAT}$, the possibility for another set of ${\nc}+24$ tile types cannot be ruled out. 
That is when the gadget patterns come into play.

	\subsection{Gadget Patterns}

Let us imagine that you are given {\nc}+24 tile types, all of which are uncolored, and asked to draw them so as for them to uniquely self-assemble $P(\phi)$ in a directed manner. 
It goes without saying that we must draw at least one tile type by each of {\nc} colors on $P(\phi)$, and 24 tile types are left uncolored. 
Applying Lemmas~\ref{lem:at_least_2-1}, \ref{lem:at_least_2-substitutor}, and \ref{lem:at_least_2-2} to $P(\phi)$ implies that we need at least 2 tile types of the 6 colors $c_{2, 1}, \ldots, c_{2, 4}$, $c_{2, 5}$ (white), and $c_{2, 6}$ (black) (see Figure~\ref{fig:tileset}).  
Now 18 extra tile types remain uncolored. 
The arguments up to now should be straightforward. 

The role of the three gadget patterns is to cooperatively force us to draw them in the same way as $T_{\rm 3SAT}$, and furthermore, allocate glues in the isomorphic manner (see Figure~\ref{fig:tileset}). 
More specifically, 
\begin{description}
\item[$P_A$:] Due to this, at least 7 of uncolored tile types are to be drawn by colors included in this pattern; 
\item[$P_B$:] Among the remaining (at most 11) uncolored tile types, at least 8 of them are to be drawn; 
\item[$P_D$:] 
	This draws the remaining (at most 3) uncolored tile types. 
	The more important role is to allocate glues onto the 12 tile types of color $c_{4, 1}, c_{4, 2}, c_{4, 3}$ as shown in Figure~\ref{fig:tileset}, and hence, implement the {\tt OR} gate, wire crossing, and {\tt XNOR} gate.  
\end{description}
At this point, the only possible coloring is to draw 4 tile types by each of $c_{4, 1}, c_{4, 2}, c_{4, 3}$ and 2 tile types by each of 15 colors $c_{2, 1}, \ldots, c_{2, 15}$ (see Figure~\ref{fig:tileset}). 

Now we explain each gadget. 
They are designed based on the parameterized mosaic pattern introduced in Section~\ref{subsec:basics}. 

	\subsubsection{Gadget Pattern $P_A$}

\begin{figure}[p]
\begin{center}
\includegraphics[scale=0.6]{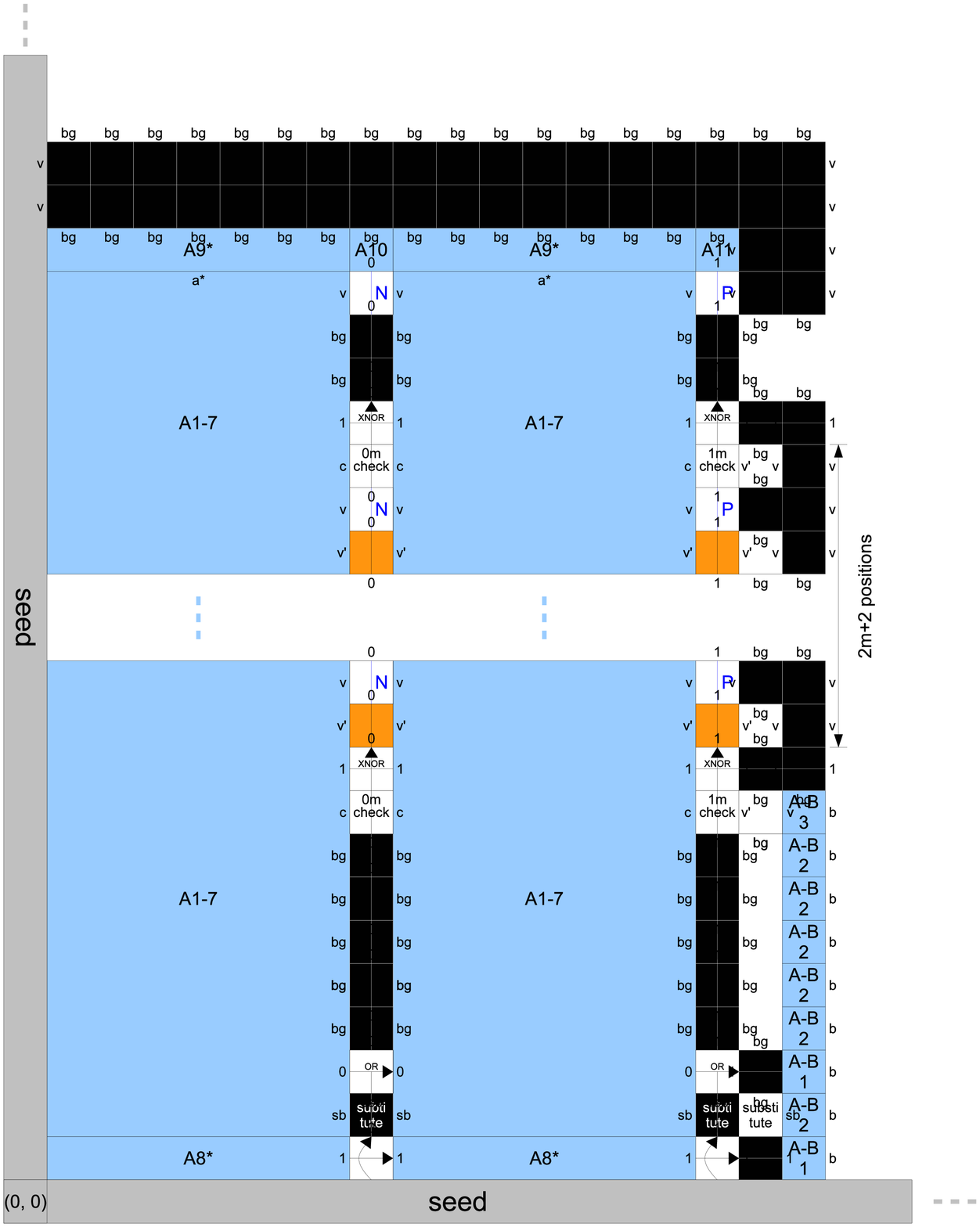}
\end{center}
\caption{The gadget pattern $P_A$.}
\label{fig:gadget_A}
\end{figure}

The gadget pattern $P_A$ is shown in Figure~\ref{fig:gadget_A}. 
This pattern employs extra 14 auxiliary colors A1-A11 and AB1-AB3 (note that these auxiliary colors will appear in $P_B$ or $P_D$, and their auxiliary colors do not appear in the other, either). 
The colors A1-A8 are used to make the mosaic pattern (in Figure~\ref{fig:gadget_A}, the mosaic is not described for the sake of clarity), which is stretched as being explained before such that to the right of $P_A$, the next gadget pattern $P_B$ can be assembled properly. 
A9-A11 seal this pattern from the top so that the variable wires $v_1, \ldots, v_m$ can carry the assignment $\vector{b}$ to the {3\sat} evaluator pattern $P_{\rm eval}(\phi)$ without trouble (see, at the top of Figure~\ref{fig:gadget_A}, we can see the lower end of the variable wire $v_m$. 
AB1-AB3 provides the glue $b$'s to the east for the gadget pattern $P_B$. 

Due to the previously-mentioned property of mosaic pattern, we need to use at least 7 colors to draw tile types that have been uncolored, and unless the 7 colors are chosen to be $c_{2, 12}, c_{2, 13}, c_{2, 14}, c_{2, 15}, c_{4, 1}, c_{4, 2}, c_{4, 3}$, one more uncolored tile type would be drawn, which is not acceptable because for $P_B, P_D$, we have to set aside 11 uncolored tile types. 
Even with the preferable (only one) choice, we further have to allocate glues to them so that 1-bit information (0/1) can be transmitted vertically because otherwise the information cannot help but penetrate the mosaic, which would cost at least one more uncolored tile type. 
It is not the case that this would allocate glues to the tile types of these 7 colors in a way isomorphic to those in Figure~\ref{fig:tileset}. 
The issue of glue allocation should be discussed after explaining the other two gadget patterns and we are convinced of the fact that the ${\nc}+24$ tile types available in total are drawn as specified in Figure~\ref{fig:tileset} (all the auxiliary colors introduced for $P_A, P_B, P_D$ are used to draw only one tile type). 

	\subsubsection{Gadget Pattern $P_B$}

\begin{figure}[tb]
\begin{center}
\includegraphics[scale=0.3]{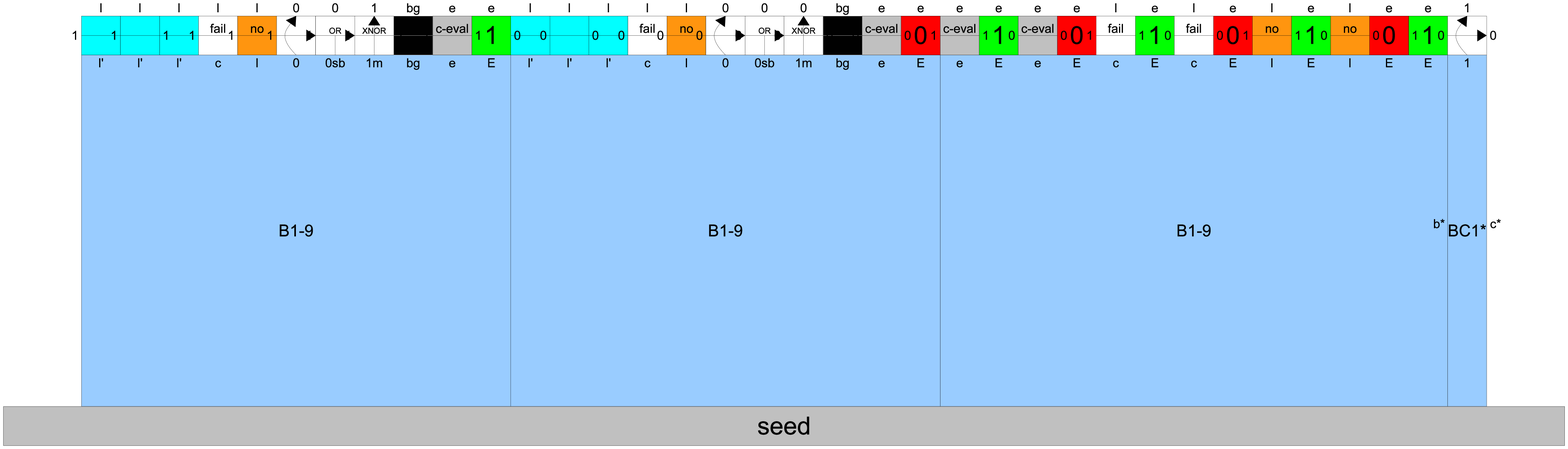}
\end{center}
\caption{The gadget pattern $P_B$.}
\label{fig:gadget_B}
\end{figure}

The gadget pattern $P_B$ is a counterpart of $P_A$ for 8 colors $c_{2, 7}, \ldots, c_{2, 11}, c_{4, 1}, c_{4, 2}, c_{4, 3}$. 
Hence, we mention only the fact that unless we use these colors to draw extra 8 tile types and allocate glues such that 0/1 signal can transmit, but rather horizontally, we would waste too many uncolored tile types. 

	\subsubsection{Gadget Pattern $P_D$}

\begin{figure}[p]
\begin{center}
\includegraphics[scale=0.6]{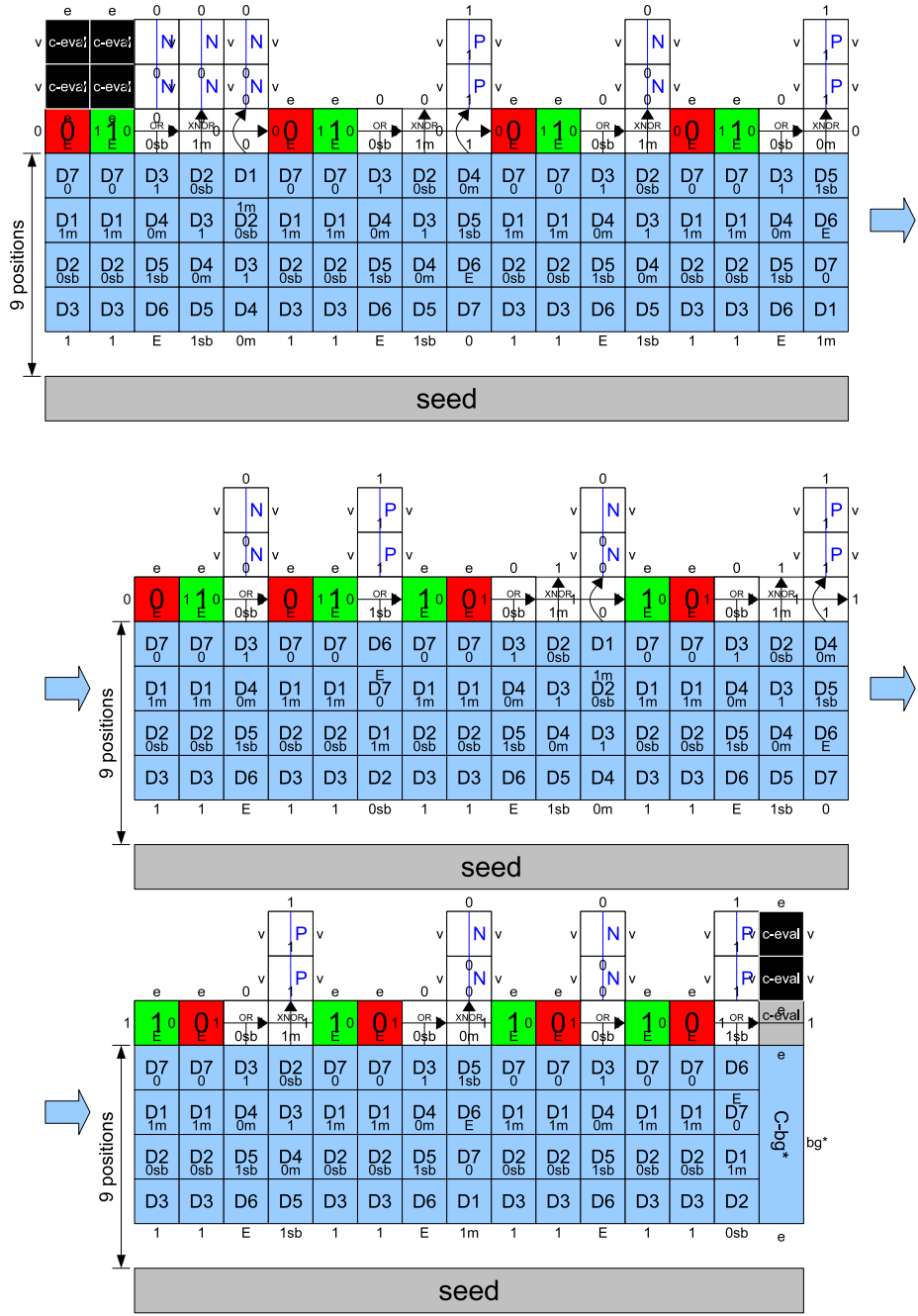}
\end{center}
\caption{The gadget pattern $P_D$.}
\label{fig:gadget_D}
\end{figure}

The gadget pattern $P_D$ is shown in Figure~\ref{fig:gadget_D}. 
This is actually a modification of the mosaic pattern using 7 colors D1-D7. 
In addition, one color C-bg is found between the border between $P_D$ and the {3\sat} evaluator pattern $P(\phi)$. 
Thus, $P_D$ introduces 8 new colors, and now all {\nc} colors have been introduced. 

We claim that this pattern needs extra 3 uncolored tile types, and they must be drawn with the colors $c_{4, 1}, c_{4, 2}, c_{4, 3}$. 
Hence, the other gadget patterns must be assembled in the intended way (with minimum consumption of the uncolored tile types). 
In summary, $c_{4, 1}, c_{4, 2}, c_{4, 3}$ each is used to draw 4 tile types (in total 12 tile types), $c_{2, 1}, \ldots, c_{2, 15}$ each is used to draw 2 tile types (in total 30 tile types), and the other 41 colors including the auxiliary ones appears on only one tile type. 

	\subsubsection{Glue Allocation}

Having completed the coloring, we briefly explain the reason why we must not only color them thus but also allocate glues to them in such a way as shown in Figure~\ref{fig:tileset}. 
Actually, it suffices to verify this isomorphism for colors that are responsible for transmitting or processing 0/1 signals for {3\sat} evaluation. 

It is helpful to observe that for a color that draws only one tile type, all tile types at positions of this color on $P(\phi)$ are identical. 
First, see the sole tile types $t_1, t_0, t_{\rm eval}$ of the respective colors $c_{1, 1}$ ({\tt 1}), $c_{1, 2}$ ({\tt 0}), and $c_{1, 4}$ ({\tt c-eval l-eval}). 
In the pattern $P(\phi)$, we find two $c_{1, 4}$-colored positions whose north neighbors are colored {\tt 1} and {\tt 0}, respectively. 
Thus, $t_1(\south) = t_0(\south) = t_{\rm eval}(\north)$. 
Then Proposition~\ref{prop:directed_RTAS_characterization} implies that $t_1(\west) \neq t_0(\west)$. 
Let $t_1(\west) = 1$ and $t_0(\west) = 0$ with $1 \neq 0$. 
Since there is a {\tt 1}-colored position whose west neighbor is colored {\tt 0}, $t_0(\east) = 1$, and likewise $t_1(\east) = 0$. 
In the same manner, for the tile types $t_P, t_N$ with respective colors $c_{1, 7}, c_{1, 8}$, we have $t_P(\west) = t_N(\west)$ but $t_P(\north) = t_P(\south) \neq t_N(\north) = t_N(\south)$. 

Now these four tile types $t_1, t_0, t_P, t_N$ can be employed to allocate glues to the other tile types. 
For example, see the right half of the gadget pattern $P_B$ (Figure~\ref{fig:gadget_B}). 
Two positions of each of the colors $c_{2, 9}, c_{2, 10}, c_{2, 11}$ are found and they are sandwiched by {\tt 1} positions and {\tt 0} positions. 
Thus, for example, we can say that there are two $c_{2, 9}$ tile types one of whose east and west glues are both {\tt 1} and the other's are both {\tt 0}. 
This guarantees that when an assignment signal (0/1) crosses the border between two {3\sat} clauses, the signal is not converted (though $P_B$ guarantees that they carry the 1-bit signal horizontally, it cannot rule out the possibility that these signals be flipping). 
$P_B$ does not contain an analogous pattern for the colors $c_{2, 7}, c_{2, 8}$, nevertheless they are also transmitting signals. 
This is because in $P(\phi)$ they always appear as a pair and hence no matter whether each of their 2 tile types converts a signal or not, it can deliver the signal correctly as long as the delivery distance is even. 

With the help of these four tile types, the gadget pattern $P_D$ allocates glues so as to implement the {\tt OR}-gate, the signal intersection, and {\tt XNOR}-gate. 
The one risk to be taken into consideration resides in the 4 tile types for the signal intersection. 
If the north glue of the D1-colored tile type is 1 and that of the D4-colored one is 0, then this converts the signal vertically. 
In order to render this conversion harmless, we encode the variable $v_i$ on the seed rather as $b_i {\tt v}^{2i} 0$; this produces a (meaningless) wire at the lower end of the variable wire $v_i$. 
As a result, any signal certainly goes through the signal intersection {\em even} times. 

	\section*{Acknowledgements}

We gratefully acknowledge valuable comments and encouragement from Ho-Lin Chen, Eugen Czeizler, David Doty, Aleck Christopher Johnson, Natasha Jonoska, Ming-Yang Kao, Steffen Kopecki, Florence Linez, Pekka Orponen, Amir Hossein Simjour, and Damien Woods. 
In particular, Ho-Lin Chen suggested encoding wires unary instead of binary, and David Doty and Damien Woods helped us to decrease the number of colors in the gadget patterns. 
The current design of {\nc}-colored pattern for the reduction was done during the first author's visit at Natasha Jonoska's research group at the University of South Florida and his stay in Nantes, France with Florence Linez. 

This research is in part financially supported by HIIT Pump Priming Project Grants 902184/T30606 to the first author.

\bibliographystyle{plain}
\bibliography{copa}

\end{document}